\documentclass{PoS}

\usepackage{amsmath}   
\usepackage{graphicx}   
\usepackage{amssymb}   
\usepackage{latexsym,cite}

\newcommand{\eq}{\begin{equation}} 
\newcommand{\en}{\end{equation}} 
\newcommand{\be}{\begin{equation}} 
\newcommand{\ee}{\end{equation}} 
\newcommand{\eqa}{\begin{eqnarray}} 
\newcommand{\ena}{\end{eqnarray}} 
\newcommand{\ba}{\begin{eqnarray}} 
\newcommand{\ea}{\end{eqnarray}}

\newcommand{\ZZ}{\hbox{{\rm Z{\hbox to 3pt{\hss\rm Z}}}}} 
 
\newcommand{\Z}{\mathbb{Z}}

\newcommand{\EQ}{\begin{equation}} 
\newcommand{\EN}{\end{equation}} 
\newcommand{\bea}{\begin{eqnarray}} 
\newcommand{\eea}{\end{eqnarray}}

\title{Integrable structures in LGTs near the deconfinement transition}

\ShortTitle{Integrable structures in LGTs near the deconfinement transition}

\author{M. Caselle, F. Gliozzi\\
        Dipartimento di Fisica Teorica dell'Universit\`a di Torino and INFN sezione di Torino,\\
        via P. Giuria 1, I-10125, Torino, Italy \\
        E-mail: \email{caselle, gliozzi@to.infn.it}}

\author{G. Delfino\\
        International School for Advanced Studies (SISSA) \\
        via Beirut 2-4, I-34014 Trieste, Italy
        and INFN sezione di Trieste\\
        E-mail: \email{delfino@sissa.it}}

\author{P. Giudice\\
        School of Mathematics, Hamilton Building Trinity College
        Dublin 2, Ireland\\
        E-mail: \email{giudice@maths.tcd.ie}}

\author{\speaker{P. Grinza} \\
        Departamento de F\'isica de Particulas and IGFAE, \\
        Universidad de Santiago de Compostela\\
        E-mail: \email{pgrinza.grinza@usc.es}}

\author{S.~Lottini\\
        Westf\"alische Wilhelms-Universität M\"unster, \\
        Wilhelm-Klemm-Strasse 9, 48149 M\"unster, Germany\\
        E-mail: \email{s.lottini@uni-muenster.de}}

\author{N. Magnoli\\
        Dipartimento di Fisica, Universit\`a di Genova and INFN sezione di Genova,\\
        via Dodecaneso 33, I-16146, Genova, Italy \\
        E-mail: \email{magnoli@ge.infn.it}}

\abstract{In this contribution we review some recent results about the emergence of 2D integrable systems in 3D Lattice Gauge Theories near the deconfinement transition. We focus on some concrete examples involving the flux tube thickness, the ratio of k-string tensions and Polyakov loops correlators in various models.}

\FullConference{The XXVII International Symposium on Lattice Field Theory - LAT2009\\
		 July 26-31 2009\\
		 Peking University, Beijing, China}

\begin{document}

\section{Introduction}
The proposal of the Svetistky-Yaffe conjecture \cite{sy82} in the early-80's came as a new interesting tool to both understand the physics of the confining/deconfining transition, and a 
smart way-out to circumvent the problem of slow Monte Carlo simulations in Lattice Gauge Theories (LGT).

In brief, the conjecture establishes a relation between the deconfining phase transition of a LGT with gauge 
group $G$ in $(d+1)$-dimensions and the symmetry breaking  phase transition of a given spin model in $(d)$-dimensions 
with symmetry group $\mathcal{C}_{G}$, i.e. the center of the group $G$. 
In particular, in the case of continuous phase transitions a standard renormalisation group 
analysis shows that the critical behaviour of the two models must be the same, i.e. they belong
to the same universality class. Hence, universal quantities like critical exponents 
and amplitude ratios can be (more easily) calculated in the spin model and then extended to the LGT.
Another key feature of the conjecture is the possibility to establish a dictionary between the observables of
the LGT and the corresponding ones in the spin model.

After more than twenty years, the Svetistky-Yaffe conjecture is considered a reliable tool to
study a LGT in the neighbourhood of the deconfinement transition. As a step forward, one can try to take advantage of the conjecture, and get a deeper quantitative understanding of a LGT in the neighbourhood of the 
deconfining transition. From this point of
view, three-dimensional LGT offers a unique opportunity, because they correspond to two-dimensional 
spin models. There are diverse motivations making such a class of spin models a privileged one.

On the one hand, they are exactly solved at the critical point, i.e. where they are relevant for the
analysis of the LGT. In fact, it was shown that a large class of spin models displaying a symmetry
breaking phase transition allows for an exact solution in terms of two-dimensional 
Conformal Field Theories (CFT) \cite{ginsp}. What makes the two-dimensional case special is the fact that
the conformal algebra becomes infinite-dimensional, and as a consequence the corresponding theory
is much more constrained that in other dimensions.
Such a richness in structure reflects at the quantitative level too. In fact, it is possible
to give a complete classification of the space of fields of a given CFT together with their
exact scaling dimension. The most direct implication of this fact is the exact calculation
of all the critical exponents of the corresponding phase transition. Another interesting
feature is given by the possibility to calculate any $n$-point correlation function
among the fields of the theory, as solutions of some well established differential equations.
In the perspective of the S-Y conjecture, such host of exact results can be directly 
extended to the corresponding LGT, giving a complete characterisation of the universality
class of the deconfinement transition. 

On the other hand, the S-Y conjecture can be used to give an effective 
description of the LGT, by means of the corresponding spin model, not only 
at the critical point but also in the scaling region.
The first non-trivial step is to show that the plaquette operator of the LGT
is mapped into a mixture of the energy and identity operators \cite{gp}.
Then, it is possible to show that the scaling region of the LGT corresponds to
the thermal perturbation of the CFT describing the critical point
of the spin model.

This aspect has important consequences, because it happens that such QFTs turn out to be
integrable for most of the cases of interest in the present context. Integrability means that an infinite number of integrals of motion exists. The main consequence in (1+1) 
dimensions is the fact that the scattering theory is
very constrained, because the $S$-matrix is factorised in products of two-body interactions, and inelastic 
processes are forbidden. 
These facts allow to write down the so-called Yang-Baxter 
equations for the 2-particle $S$-matrix. Then, such an $S$-matrix can be computed exactly by imposing the 
previous equations and the usual requirements of unitarity and crossing (for a review about Integrable QFTs 
see \cite{revgius}). 
An obvious consequence is that also the spectrum of the masses of the bound states of the theory is known 
exactly, since they are represented by the simple poles of the $S$-matrix in the physical strip.

Another useful aspect of dealing with an integrable theory is the possibility to use the spectral
expansion (form factors can be computed exactly in integrable QFTs) for correlation functions. 
Such a fact gives the possibility to
describe with a good deal of accuracy the large distance behaviour of the correlation functions. 
As we will see in the following, this is a key point in our approach. 

In the present contribution we gathered our main results obtained in several 
models, and with different types of observables. In sect.~2 we
report on the study of the baryon (three-quark) potential in the 3D SU(3) LGT using the three-spin correlation
function of the 2D 3-state Potts model \cite{cdgmj}. Sect.~3 is devoted to the study of the behaviour of the flux tube thickness 
in the 2D SU(2) LGT via the corresponding 2D Ising model \cite{cgmflux}. Finally, in sect.~4 we report on the exact
calculation of the ratio of k-string tensions in the 3D $Z_4$ LGT my means of the mapping on the 2D
Sine-gordon model \cite{cgggl}.
 
\section{Three-quark potential \cite{cdgmj}} 
In these last years much interest has been attracted by the study of the three-quark  
potential in LGT. Besides the obvious   
phenomenological interest of the problem, the three-quark potential is also a   
perfect tool for testing our understanding of the flux tube model of confinement   
and of its theoretical description in terms of effective string models.
Thanks to the improvement in lattice simulations (a summary of numerical   
results can be found in~\cite{aft02}), the qualitative   
behaviour of the three-quark potential is now rather well understood (for a   
recent review see~\cite{fj05}). Let us briefly review the key points.

For large interquark distances the three-quark   
potential is well described by the so-called {\bf Y} law which assumes    
a flux tube configuration composed by    
three strings which originate from the three quarks and join in the    
Steiner point which has the property of minimising the overall length of the   
three strings. This picture is also in agreement with what one would naively   
find using standard strong coupling expansion. Notice however that due to the   
roughening transition this is only a qualitative indication, and cannot be   
advocated as a ``proof'' of the {\bf Y} law.

At shorter distances a smooth crossover toward the so called {\bf $\Delta$}   
law is observed. According to the {\bf $\Delta$} law the three-quark    
potential is well approximated by the sum of the three two-quark interactions. 
More precisely the {\bf $\Delta$} law assumes that the three-quark correlator  
(let us call it $G_3(x_1,x_2,x_3)$ where $x_j$ denotes the position of the $j^{th}$ quark) 
is related to the quark-antiquark correlator $G_2(x_i,x_j)$ as follows: 
\eq 
G_3(x_1,x_2,x_3)\sim \sqrt{G_2(x_1,x_2)~G_2(x_2,x_3)~G_2(x_1,x_3)} 
\label{delta} 
\en    
thus leading to  a potential which increases linearly with the sum of the three   
interquark distances. The scale where the transition between these two   
behaviours seems to occur, according to the most recent simulations, is around   
0.8 fm.   

To improve our understanding of the baryon states,    
it would be important now to have some quantitative insight in the above    
described picture, as well as to have some theoretical argument to explain    
why instead of having a single shape stable for all the interquark distances a   
$\Delta \to Y$ crossover occurs. Moreover, since the crossover region happens to   
occur exactly in the range of distances which is interesting from a   
phenomenological point of view, it would be important to have some kind of   
theoretical description of this crossover with which to compare the numerical data.   
   
In this respect the present    
study of the three-point function in the 2d ${\mathbb{Z}}_3$ Potts   
model is a perfect laboratory to address this problem. Besides the obvious   
similarity  of the two settings it is also possible to find a direct relation,   
since by the Svetitsky-Yaffe conjecture the behaviour at high temperature of the three-quark  
correlator for a $SU(3)$ or a ${\mathbb{Z}}_3$ gauge model in (2+1)   
dimensions is mapped into the behaviour of the 2d ${\mathbb{Z}}_3$ Potts three-point   
function, in analogy to what happens for the quark-antiquark potential which is   
mapped onto the $\langle\sigma \bar\sigma\rangle$ correlator.

In the large distance regime, the form factor approach (spectral expansion) plays a central
role due to the integrability of the theory. The appearance of the Steiner
point is quite natural in such a framework (even if completely non-trivial), and it is 
strongly related to the integrable structure of the model. Here we quote
the final result (leading and sub-leading orders) in the case of an equilateral 
triangle of side $R$ (see \cite{cdgmj} for details)
\bea   
G^{(3)} (x_1,x_2,x_3)  \ \simeq \   &&     
\frac{(F_{\bar A}^\sigma)^3 \; \Gamma^{\bar A}_{AA} }{ \pi} K_0   
\left(\sqrt{3}\,  m R  \right)+      
\nonumber  \\   
&&  + \    
6 (F_{\bar A}^\sigma)^2 \int_{-\infty}^{\infty} \frac{\textrm{d} \theta}{(2 \pi)^2}    
F_{ A  A}^{ \sigma} ( \theta +  i \pi/ 3) K_0 \left( 2 m R \cosh \frac{\theta}{2}   
\right)   
\label{g3equi}   
\eea    
where $r_Y =\sqrt{3}\,  R $ is the minimal distance given by the emergence of the
Steiner point. 
On the other hand, the short distance regime is studied by means of the so-called
Conformal Perturbation Theory (CPT) \cite{CPT}, which is a perturbative approach built upon the
exact CFT solution of the model at the critical point (integrability is not needed here). 
We first notice that, at
criticality, the {\bf $\Delta$} law is exact. 
In fact we see that the 
relation (\ref{delta}) is fulfilled exactly by 2- and 3-p correlator 
functions of a given CFT. Hence, we used the CPT framework to compute the corrections
induced by the thermal perturbation, and we are able to explicitly compute
the perturbative corrections to the {\bf $\Delta$} law.

\begin{figure}[htbp]   
  \centering   
  \includegraphics[width=8cm]{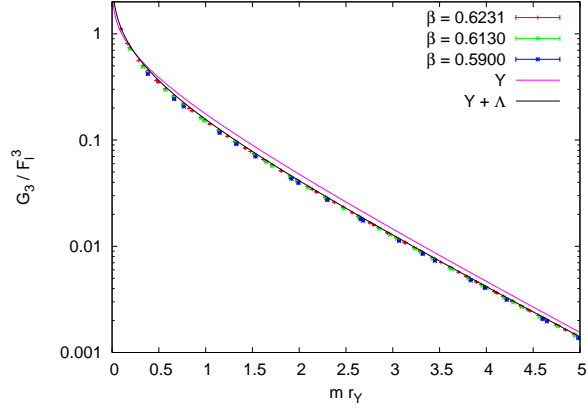}   
  \caption{Three-point function (suitably normalised, see \cite{cdgmj} for details): spectral expansion (form factors), and Monte Carlo data.} 
  \label{fig:3-point}   
\end{figure}   

\begin{figure}[tbh] 
  \centering 
  \includegraphics[width=8cm]{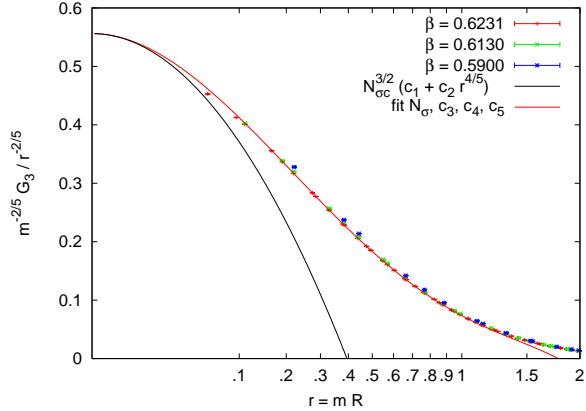} 
  \caption{Three-point function (suitably normalised, see \cite{cdgmj} for details): short-distance expansion (Conformal Perturbation Theory), and Monte Carlo data.} 
  \label{fig:tpshort} 
\end{figure} 

From a physical point of view the scenario which emerges 
 is a smooth crossover, as the distance among the three points increases,   
from a short distance behaviour in which the three point function is   
dominated by the three spin-spin interactions  along the edges of the triangle to   
a large distance behaviour in which the strong coupling expectation (the three   
spins joined by a path of minimal length) is fully realised.   
   
This scenario is confirmed by the numerical simulations which turn out to be in   
remarkable agreement with our theoretical results (see figures \ref{fig:3-point}, \ref{fig:tpshort}).      
Indeed   
the important consequence of having exact analytic results    
for the two expansions is that we    
are able to compare our predictions with triangles of any size, both smaller and larger than the correlation length.

In conclusion, we would like to stress that the S-Y conjecture allowed us to check with some solid
field-theoretic argument that, at least near the deconfinement temperature, a crossover between  {\bf Y} and  {\bf $\Delta$} laws occurs at some intermediate scales.

\section{Flux tube thickness \cite{cgmflux}} 
The distinctive feature of the interquark potential in a confining gauge theory is that the  
colour flux is confined into a thin flux tube, joining the quark-antiquark pair. As it is well known 
the 
quantum fluctuations of this flux tube, which are assumed to be described by a suitable effective string model, 
lead to a logarithmic increase of the width of the flux tube as a function of the interquark distance $R$. This 
behaviour was discussed many years ago by L\"uscher, M\"unster and Weisz in~\cite{lmw80} and is one of the 
most stringent predictions of the effective string description of confining LGTs.  

A natural question is what happens of this picture at the deconfinement point. One would naively expect a sudden 
jump of the flux tube thickness from a log to a linear dependence from the interquark distance. However we shall 
show in this work that this is a 
misleading picture.

A tentative answer to this question can be obtained in the effective string framework.  
By using a duality transformation it is possible to show that as the temperature 
increases the log behaviour smoothly moves to a linear behaviour, thus excluding a log to linear transition at 
the deconfinement point. 
The simultaneous dependence of the flux tube thickness on the two variables $R$ and $N_t$ 
 can be evaluated exactly only 
in the gaussian limit\footnote{Including higher order terms in the effective string action makes the problem too 
difficult, even if some recent result in the framework of the covariant quantisation suggest that some 
simplification could occur if one chooses to study the whole Nambu-Goto action \cite{bc05bcf06}.}. 
For the details of the calculations we refer the reader to the paper~\cite{cgmv95}.

To the purpose of the present work, we are only interested in the two asymptotic limits: large $N_t$ and finite $R$  
(which is the zero 
temperature limit where we expect a log type behaviour) and the opposite one: large $R$ and small $N_t$ which is 
high temperature limit.  
 
One finds: 
\eq 
w^2\sim\frac{1}{2\pi\sigma}\log(\frac{R}{R_c})~~~~~~~~~~~(N_t>>R>>0)  
\en 
 
\eq 
w^2\sim\frac{1}{2\pi\sigma}(\frac{\pi R}{2 N_t}+\log(\frac{N_t}{2\pi}))~~~~~~~~~~~(R>>N_t)  
\label{espred} 
\en 
As it is easy to see in the second limit the logarithmic dependence is on $N_t$  
(the inverse of the temperature) and 
not on $R$ which appears instead in the linear correction. 
However this result strongly relies on the effective string approximation (even worse 
on the gaussian limit of the effective string) and it would 
be nice to have some kind of independent evidence. 

Hence, we propose an alternative way to address the above question in the vicinity of the deconfinement 
transition using the Svetitsky-Yaffe conjecture which is a very powerful tool  
to study the finite T behaviour of a confining LGT in the vicinity of the deconfinement point, at least   
for those LGTs whose  deconfinement transition is of second order.

This gives us a non trivial opportunity to check the effective string predictions.  
If we choose a (2+1) dimensional LGT with a 
gauge group with center $Z_2$ (like the gauge Ising model or the SU(2) or SP(2) LGTs which all have continuous 
deconfinement transitions), the target spin model is the 2d Ising model in the high temperature symmetric phase 
for which several exact results are known. In particular we shall see that it is possible to study analytically 
the equivalent of the flux tube thickness.
Leaving the details to \cite{cgmflux}, it is possible to show that, by means of the spectral expansion over form factors, for very large separations between the spins (quarks) the flux tube thickness in the Ising model behaves like
\bea 
w^2 \ \simeq \  \frac{\pi}{2} \, \frac{R}{\sigma N_t} + \dots, \ \ \ \ \ \ \  N_t \to \beta_c. 
\label{result} 
\eea  

Remarkably enough the results that we find agree, up to non-universal constants, with the effective 
string ones thus strongly supporting the idea of a smooth transition from a log to a linear behaviour as the 
temperature increases.  
A numerical confirmation of such a scenario has been recently found in \cite{ac08}.

\section{K-string tensions ratios  \cite{cgggl}} 

In a recent paper \cite{confanomaly}, it was argued from simple 
scaling properties of suitable Polyakov loop correlators that the k-string 
tensions have the following 
low temperature asymptotic expansion 
\EQ
\label{tension}
	\sigma_k(T) = \sigma_k - c\frac{\pi}{6}T^2 + \mathcal{O}(T^3) \; ; \; 
	c = (d-2)\frac{\sigma_k}{\sigma	} \; ,
\EN
where $c$ is the central charge of the underlying 2D conformal field theory 
describing the IR behaviour of the k-string. As a consequence, their ratios are expected to be constant up to 
 $T^3$ terms:
\EQ
\frac{\sigma_k(T)}{\sigma(T)}=
\frac{\sigma_k}{\sigma} + \mathcal{O}(T^3) \,\,.
\label{ratios}
\EN   

The low temperature data presented in support of this expectation were 
taken from Monte Carlo simulations on a particular system, namely a 
(2+1)-dimensional $\Z_4$ gauge model, which is the simplest 
exhibiting more than just the fundamental string. 

The main conjecture we want to verify in this work is that 
$\sigma_k(T)/\sigma(T)$, at least in that $\Z_4$ gauge system, is in fact 
independent of the temperature in the \emph{whole} of the confining regime. 
To check this idea, a handy fact comes useful, namely that,
as the system approaches the deconfinement transition, and the 
string picture begins fading, another approach is made available by the 
Svetitsky-Yaffe (S-Y) conjecture \cite{sy82}, which allows to reformulate the 
system in a totally different perspective, based on a two-dimensional 
integrable theory 
in which, however, the near-$T_c$ counterpart of the low-temperature 
result cited above can be nicely found.
It turns out that the deconfinement transition of the 3D 
$\Z_4$ gauge model is second order and, according to the SY conjecture,
  belongs to the same universality class of the 2D symmetric 
Ashkin-Teller (AT) model. As a matter of fact, such a model possesses a 
whole line of critical points along which the critical exponents vary 
continuously. The S-Y conjecture tells us that if a 2+1 dimensional gauge 
model with center $\Z_4$ displays a second-order transition, then its universality class is associated to a 
suitable point of the critical line of the 
2D AT model \cite{baxterbook}. For instance, it has been argued \cite{su4deforcrand} that the critical 2+1 $SU(4)$ gauge theory 
belongs to the universality class of a 
special point of the AT model, known as the  four-state Potts 
model. More generally, the class of models with gauge group $\Z_4$ depends 
on two coupling constants $\alpha$ and $\beta$, and the universality 
class of the deconfining point $P$ varies with the ratio $\alpha/\beta$.

The two-dimensional AT model can
be seen in the continuum limit as a bosonic conformal field theory plus a 
massive perturbation  (i.~e.~a Sine-Gordon theory)
driving the system away from the critical line. 
Thus, a map between (a neighbourhood of) the AT critical line 
and the Sine-Gordon phase space is provided.
This theory is integrable, and the masses of its lightest physical states 
(first soliton and first breather mode, of masses $M$ and $M_1$ \cite{refSG}) 
correspond to the tensions 
$\sigma(T)$ and $\sigma_2(T)$ near $T_c$, whose ratio
can be analytically 
evaluated and turns out to be
\eq
\lim_{T\to T_c}\frac{\sigma_2(T)}{\sigma(T)}=
\frac{M_1}{M}=2\sin\frac\pi2(2\nu-1)\;\;,
\label{main}
\en
where $\nu$ is the thermal exponent in two dimensions.

As a consequence, on the gauge side, we have two different ways to verify the conjecture. One is to  directly 
estimate the ratio $M_1/M$ by measuring the Polyakov-Polyakov correlators in the two non-trivial representations
 of $\Z_4$ near the deconfining temperature.
The other is to evaluate the thermal exponent of the gauge system at the
deconfining temperature. Either method gives a value of ${M_1}/{M}$ which 
can be compared with the ratio $\sigma_2/\sigma$ evaluated at $T=0$.
In particular, choosing $(\alpha,\beta)=(0.050,0.207)$, we obtained the following results:
\begin{itemize}
\item{at $T=0$ we have \cite{confanomaly}
\EQ
	\frac{\sigma_2}{\sigma} = 1.610(13) \; ,
\label{ratiosigma}
\EN
}
\item{at $T \to T_c^{-}$, the fit of the data for the Polyakov-Polyakov correlators in the fundamental and double-fundamental reps gives \cite{cgggl}
\eq
\sigma_2(T\sim T_c)/\sigma(T\sim T_c)=\frac{M_1}{M}=1.612(46)\,,
\label{mffovermf}
\en

}
\item{at $T \to T_c^{-}$, the evaluation of the thermal critical exponent $\nu$ gives \cite{cgggl} (the systematic error is quoted in square brackets)
\EQ
\sigma_2(T\sim T_c)/\sigma(T\sim T_c)=\frac{M_1}{M} = 2\sin\frac\pi2(2\nu-1)= 1.6124(71)[102] \; ,
\label{mfromnu}
\EN
}
\end{itemize}
where we can see that the latter estimates at $T \to T_c^-$ give compatible results which nicely agree with the ratio 
$\sigma_2/\sigma$ evaluated at $T=0$.

{\bf Acknowledgements} The work of P.G. is partially supported by MEC-FEDER (grant FPA2008-01838), by the Spanish Consolider-Ingenio 2010 Programme CPAN (CSD2007-00042) and by Xunta de Galicia (Conselleria de Educacion and grant PGIDIT06PXIB296182PR).

\end{document}